\documentclass[prb,aps,twocolumn,showpacs,floats]{revtex4}
\usepackage{graphicx,amsmath,epsf}
\usepackage{subfigure}

\newcommand{\vn}{\vec n}

\newcommand{\vs}{\vec s}

\newcommand{\vT}{\vec T}

\renewcommand{\vec}[1]{\mathbf{#1}}

\newcommand{\vm}{\mathbf{m}}

\begin{document}

\title{Spin torque from tunneling through impurities in a magnetic
  tunnel junction}

\author{Turan Birol and Piet W.\ Brouwer}

\affiliation{Laboratory of Atomic and Solid State Physics, Cornell
  University, Ithaca, NY 14853, USA}

\date{\today}


\begin{abstract}
We calculate the contribution to the spin transfer torque from
sequential tunneling through impurities in a
magnetic tunnel junction. For a junction with weakly polarized
ferromagnetic contacts, the torque is found to be in the plane spanned
by the magnetizations of the ferromagnetic contacts and proportional to
$\sin \theta$, where $\theta$ is the angle between the magnetic
moments. If the polarization is larger, the torque acquires a
significant out-of-plane component and a different dependence on $\theta$.
\end{abstract}

\maketitle

\section{Introduction}
A magnetic tunnel junction consists of two ferromagnetic layers,
separated by an insulator. \cite{kn:zhang2003,kn:tsymbal2003b}
Magnetic tunnel junctions play a prominent role in proposals
for magnetic memory applications, where the information is stored in
the relative orientation of the magnetizations of the two
ferromagnetic layers. The two essential operations for the realization
of a memory element, reading and writing, rely on the tunneling
magnetoresistance effect (the observation
that the resistance of the junction depends on the relative
orientation of the two ferromagnets \cite{kn:julliere1975})
and the spin transfer torque (the fact
that passing a large current through a
magnetic tunnel junction exerts a torque on the magnetizations of the
two layers. \cite{kn:slonczewski1996,kn:berger1996})
%
Although both effects also exist in metallic magnetic
junctions, tunneling junctions have a significantly higher
magnetoresistance and an impedance that is better matched to the
requirements of semiconductor technology. \cite{kn:parkin1999}

Although the tunneling magnetoresistance effect has been studied
experimentally and theoretically for more than 30 years, the spin
transfer effect in magnetic tunnel junctions had not received
attention until recently. The spin transfer torque consists of two
contributions, a component in the plane spanned by the magnetization
directions of the two ferromagnetic layers in the junction, and a
component transverse to that plane. The out-of-plane component is
also referred to as the ``field-like'' torque or the
``nonequilibrium exchange interaction''. The in-plane torque
component can be understood in terms of the non-conservation of spin
for electrons tunneling between the ferromagnetic
layers. \cite{kn:slonczewski2005} The out-of-plane torque component
has been attributed to a more subtle interference between electrons
reflected off the front and back ends of the insulating
barrier. \cite{kn:xiao2008,kn:manchon2008} Theoretical predictions
exist for both torque contributions, with similar conclusions based
on tight-binding calculations that include the full band structure
of the
junction \cite{kn:heiliger2008,kn:theodonis2006,kn:kalitsov2006} and
free electron models that neglect most band structure
effects. \cite{kn:xiao2008,kn:manchon2007,kn:manchon2008} On the
experimental side, quantitative measurements of both components of the
spin transfer torque in tunnel junctions are possible using
spin-transfer-driven ferromagnetic resonance.
\cite{kn:tulapurkar2005,kn:sankey2008}

Whereas most of the theoretical approaches to the spin-transfer torque
in magnetic tunnel junctions address ideal junctions, impurities and
defects are known to be abundant in the MgO magnetic tunnel junctions
used in experiments. \cite{kn:mather2006,kn:cha2007,kn:miao2008}
While impurities inside the barrier
are known, both theoretically \cite{kn:tsymbal2003,kn:velev2007b} and
experimentally, \cite{kn:jansen1999,kn:miao2008} to strongly
affect tunneling magnetoresistance, there has been less attention to
their effect on the spin transfer torque.
\cite{kn:manchon2007,kn:zhuravlev2005}

Unlike for a theory of spin torque and magnetoconductance in ideal
junctions, where both phenomena are attributed to electron states that
are extended on both sides of the junction, impurity-mediated
transport involves localized states inside the
barrier. Inside the barrier, electron-electron interactions are poorly
screened, and interactions cannot simply be accounted for on the
mean-field level. In this letter,
we describe the impurity-mediated transport using a
model of sequential tunneling that is able to incorporate interaction
effects exactly. The sequential tunneling picture is valid for high
temperatures or voltages (temperature
and/or voltage larger than the width of the impurity state), which
makes our theory applicable to all but the thinnest junctions. This
puts our theory in an entirely different parameter range than
Refs. \onlinecite{kn:zhuravlev2005} and \onlinecite{kn:manchon2007}, which addressed zero
temperature torques in the absence of interactions.

The concept of sequential tunneling has been used previously to
describe spin-dependent transport through metal nanoparticles (or
``quantum dots'') tunnel coupled to ferromagnetic
electrodes.
\cite{kn:waintal2003,kn:cottet2004,kn:koenig2003,kn:braun2004}
For
these systems, one considers how a current between the
ferromagnetic contacts affects the spin accumulated on the
nanoparticle.
In the present context, we are interested in the opposite effect, the
back-action of the current on the polarization of the ferromagnetic
contacts.
Below, we first analyze the current $I$ and torques $\vT_L$ and
$\vT_{R}$ on the two ferromagnets in the tunnel junction for
sequential tunneling through a single impurity with a single (spin
degenerate) energy level, following ideas laid out in the theory of
spin-dependent transport through quantum dots. We then turn to the
case of magnetic tunnel junctions, where impurity-mediated
contributions to the current and torque have to be summed over many
impurities in the junction.
%

\section{Sequential tunneling through a single impurity} 
\label{sec:2}

\begin{figure}
\centering{}
\includegraphics[width=0.95\hsize]{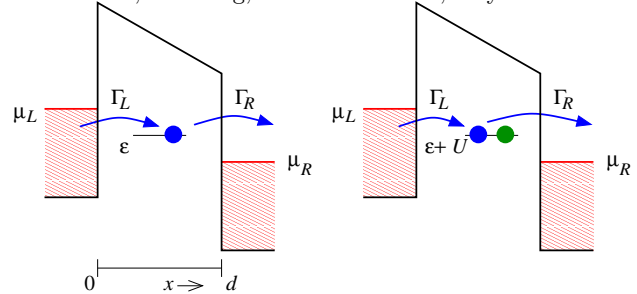}
\caption{Schematic picture showing the impurity energy level and the
  chemical potentials of the ferromagnetic reservoirs. The left panel
  is for the case $\mu_R < \varepsilon < \mu_L$, in which the
  occupation of the impurity site fluctuates between $0$ and $1$. The
  right panel is for the case $\mu_R < \varepsilon + U < \mu_L$, in
  which the impurity site occupation varies between $1$ and $2$. These
  two cases are labeled I and II in the main text.
  \label{fig:setup}}
\end{figure}

We denote the
energy of the singly-occupied and doubly-occupied impurity level by
$\varepsilon$ and $2 \varepsilon + U$, respectively, where the
interaction energy $U$ accounts for the Coulomb repulsion of the two
electrons in a doubly-occupied impurity site. The left (L) and right
(R) ferromagnets are held at chemical potential $\mu_{L}$ and
$\mu_{R}$, respectively. For definiteness, we assume that the bias
$eV = \mu_L - \mu_R > 0$. (See Fig.\ \ref{fig:setup} for a schematic
picture.)
The direction of the magnetization of each
ferromagnetic contact is denoted by the unit vector
$\vec{m}_{\alpha}$, $\alpha=L,R$. Tunneling between the impurity
level and the reservoirs is described by tunneling rates
$\Gamma_{L\sigma}$ and
$\Gamma_{R\sigma}$, where $\sigma=\uparrow,\downarrow$ for the
majority and minority spin directions, respectively.

The theory of sequential tunneling is applicable if the temperature
$T$ is much larger than the tunneling rates
$\Gamma_{\alpha\sigma}$. In this regime, interference
effects (such as those responsible for an impurity-mediated zero-bias
torque \cite{kn:zhuravlev2005})
are smeared out and impurity-mediated transport can be
described using the probabilities $p_j$ that the impurity
level is occupied by $j$ electrons, $j=0,1,2$, and the
expectation value $\vec{s}$ of the impurity
spin. \cite{kn:braun2004} Their time derivatives
$\dot{p_j}$ and $\dot{\vs}$ are expressed in terms of particle and
spin currents onto the impurity site,
\begin{eqnarray}
  \dot{p_0} &=& - I_{L}' - I_{R}', \nonumber \\
  \dot{p_1} &=& I_{L}' + I_{R}' - I_{L}'' - I_{R}'',\ \
  \dot{\vs} = \vec{I}_{L} + \vec{I}_{R},
  \nonumber \\
  \dot{p_2} &=& I_{L}'' + I_{R}'',
  \label{eq:peq}
\end{eqnarray}
where $I_{\alpha}'$ and $I_{\alpha}''$ are the currents from contact
$\alpha$
for processes in which the occupation
changes from $0$ to $1$ and from $1$ to $2$, respectively, and
$\vec{I}_{\alpha}$ is the spin current onto the impurity site, $\alpha
= L,R$.
The component of
$\vec{I}_{\alpha}$ perpendicular to $\vm_{\alpha}$ causes
a torque $\vec{T}_{\alpha}$ \cite{kn:slonczewski1996},
\begin{equation}
  \vec{T}_{\alpha} = \hbar (\vec{I}_{\alpha} \times \vm_{\alpha})
  \times \vm_{\alpha}.
\end{equation}
The torques are
decomposed into an in-plane
component $T_{\alpha \parallel}$, $\alpha=L,R$, and an
out-of-plane component $T_{\perp}$,
\begin{eqnarray}
  \vT_L &=& T_{L\parallel} \vm_{L} \times \vn
  + T_{\perp} \vn, \label{eq:TL} \\
  \vT_R &=& T_{R\parallel} \vm_{R} \times \vn
  - T_{\perp} \vn \label{eq:TR},
\end{eqnarray}
where $\vn$ is the unit vector perpendicular to $\vm_L$ and
 $\vm_R$,
\begin{eqnarray}
  \vn &=& (\vm_L \times \vm_R)/|\vm_L \times \vm_R|.
\end{eqnarray}

Following Ref.\ \onlinecite{kn:braun2004}, the
currents $I_{\alpha}'$, $I_{\alpha}''$, and $\vec{I}_{\alpha}$
are expressed in terms of the tunneling rates
$\Gamma_{\alpha \sigma}$ and the distribution functions
$f_{\alpha}(\xi) = [1 + \exp{((\xi-\mu_{\alpha})/T)}]^{-1}$
of the two reservoirs,
\begin{widetext}
\begin{eqnarray}
  \label{eq:rate}
  I_{\alpha}' &=& 2 \Gamma_{\alpha+} p_0 f_{\alpha}(\varepsilon)
  \nonumber \mbox{}
  - [1 - f_{\alpha}(\varepsilon)]
  \left(\Gamma_{\alpha+} p_1
  + 2 \Gamma_{\alpha-} \vm_{\alpha} \cdot \vs \right), \nonumber \\
  I_{\alpha}'' &=&
  f_{\alpha}(\varepsilon+U)
  \left(\Gamma_{\alpha+} -
  2 \Gamma_{\alpha-} \vm_{\alpha} \cdot \vs \right)
  \nonumber \mbox{}
  - 2 \Gamma_{\alpha+} p_2 [1 - f_{\alpha}(\varepsilon+U)], \nonumber \\
  \vec{I}_{\alpha} &=&  h_{\alpha}
  \vec{m}_{\alpha} \times \vec{s}
  + \Gamma_{\alpha-} f_{\alpha}(\varepsilon) p_0 \vm_{\alpha}
  \nonumber  \mbox{}
  - \Gamma_{\alpha-} [1 - f_{\alpha}(\varepsilon + U)] p_2
  \vec{m}_{\alpha}
  \nonumber \\ && \mbox{}
  - [1-f_\alpha(\varepsilon)]
  \left( \Gamma_{\alpha+} \vs +
  \Gamma_{\alpha-} p_1 \vm_{\alpha}/2 \right)
  \nonumber  \mbox{}
  - f_{\alpha}(\varepsilon+U)
  \left(\Gamma_{\alpha+} \vs -
  \Gamma_{\alpha-} p_1 \vm_{\alpha}/2 \right).
\end{eqnarray}
\end{widetext}
Here $\Gamma_{\alpha\pm} = (\Gamma_{\alpha\uparrow} \pm
\Gamma_{\alpha\downarrow})/2$, and
\begin{eqnarray}
  h_{\alpha} &=& \frac{1}{2\pi} {\rm P} \int_{-\infty}^\infty d\xi
    \frac{U(1 - 2 f_{\alpha}(\xi))
  \Gamma_{\alpha-}}
  {(\varepsilon - \xi) (\varepsilon + U - \xi)}
  \nonumber \\ & \approx &
  \frac{\Gamma_{\alpha-}}{\pi} \ln \left| \frac{\varepsilon - \mu_{\alpha}}{\varepsilon + U
    - \mu_{\alpha}}\right|
  \label{eq:hdef}
\end{eqnarray}
is an effective exchange field arising from virtual tunneling from the
impurity level to the reservoirs. \cite{kn:koenig2003} The
exchange field is an interaction effect: $h_{\alpha}=0$ in the absence
of Coulomb repulsion on the impurity site.

Sequential tunneling takes place if at least one of the energies
$\varepsilon$ and $\varepsilon + U$ lies between $\mu_L$ and
$\mu_R$, see Fig.\ \ref{fig:setup}.
Since typical interaction energies $U$ are large (up to several
$eV$), we assume that $eV \ll U$. In this regime, the occupation of
the impurity site can change by at most one electron. Depending on
whether $\varepsilon$ or $\varepsilon+U$ lies between the chemical
potentials of the source and drain contacts, the occupation of the
impurity site fluctuations between $0$ and $1$ (``case I'') or $1$
and $2$ (``case II''). In case I, the only nonzero charge currents
are $I_L'$ and $I_R'$, whereas in case II, the nonzero charge
currents are $I_L''$ and $I_R''$. Below, we give the relevant
expressions for case I only. Case II follows from the expressions
below by making the substitutions $L \leftrightarrow R$ and $h
\leftrightarrow - h$.

Solving Eqs.\ (\ref{eq:peq})--(\ref{eq:rate}) for $T \ll eV$,
we then calculate the current $I = I_L$ through the impurity,
as well as the torques $\vT_L$ and $\vT_R$.
We then find
\begin{widetext}
\begin{eqnarray}
   I &=& D^{-1} \Gamma_{L+} \left[
  (\Gamma_{R+}^2 + h_L^2 + h_R^2 + 2 h_L h_R \cos \theta)
  (\Gamma_{R-}^2 - \Gamma_{R+}^2)
  -
  \Gamma_{R-}^2 h_L^2 \sin^2 \theta \right], \\
  T_{R\parallel} &=&
  2\hbar D^{-1} \sin \theta \left\{
  \Gamma_{L-} (\Gamma_{R-}^2 - \Gamma_{R+}^2)
  [\Gamma_{R+}^2 + (h_R^2+h_L^2+h_L h_R \cos \theta)]
  - h_L^2 \Gamma_{R-} (\Gamma_{R-} \Gamma_{L-} - \Gamma_{L+}
  \Gamma_{R+} \cos \theta) \right\},
  \label{eq:trp} \\
  T_{L\parallel} &=&  2\hbar D^{-1} h_L  \sin \theta \left[
  \Gamma_{L-} \Gamma_{R-}^2(h_R + h_L \cos
  \theta)
  - \Gamma_{R+} (h_L \Gamma_{L+} \Gamma_{R-} + h_R \Gamma_{R+}
  \Gamma_{L-}) \right],\label{eq:tlp} \\
  T_{\perp} &=& 2\hbar
  D^{-1} h_L  \sin \theta \left\{ \Gamma_{L-} \Gamma_{R+} h_L h_R - \Gamma_{L+}
  \Gamma_{R-} (\Gamma_{R+}^2 + h_R^2) -
  [\Gamma_{L+} \Gamma_{R-} h_L h_R -
  \Gamma_{L-} \Gamma_{R+} (\Gamma_{R-}^2 + h_R^2) ] \cos \theta
  \right\}, ~~~~ \label{eq:tp}
\end{eqnarray}
with
\begin{eqnarray}
  D &=
  2(\Gamma_{R+}^2 + h_L^2 + 2 h_L h_R \cos \theta + h_R^2)
  [\Gamma_{R-}(\Gamma_{R-}+&2\Gamma_{L-} \cos \theta) - \Gamma_{R+}( \Gamma_{R+}+ 2 \Gamma_{L+})]
  \nonumber \\ && \mbox{}
  + 2\Gamma_{R-} h_L (2 \Gamma_{L-} h_R - \Gamma_{R-} h_L )
  \sin^2 \theta.
\end{eqnarray}
\end{widetext}

It is important to point out that the in-plane torque components $T_{R\parallel}$ and $T_{L\parallel}$ on the magnetizations of the source and drain contacts are, in general, not equal. (The out-of-plane components are equal because of conservation of angular momentum.) The origin of this effect is the Coulomb interaction on the impurity site, which lifts the symmetry between source and drain reservoirs. To see this, recall that in case I (for which the above equations are derived), the Coulomb interacation forbids double occupancy of the impurity site. Since the electron spin is not changed when an electron tunnels onto an empty impurity level, the electron spin may be changed only when the electron tunnels off the impurity site, or because of the action of the exchange fields $h_L$ and $h_R$. This explains why $T_{R\parallel}$ is nonzero to lowest order in the exchange fields, whereas $T_{L\parallel}$ and $T_{\perp}$ are of higher order in $h_L$ and $h_R$, respectively. 

\section{Impurity-mediated tunneling in a magnetic tunnel junction} 

In
a magnetic tunnel junction there will be many impurities,
distributed spatially throughout the insulating spacer layer, and
with a distribution of energy levels $\varepsilon$. (The energy
$\varepsilon$ depends on the electrostatic environment of the
impurity, hence its fluctuations.) In order to add the contributions
from all impurities, we characterize the impurity configuration by
the density $\rho(\varepsilon, x)$ of impurities per volume and per
energy, where $0 < x < d$ is the distance from the source reservoir,
$d$ being the width of the insulating spacer layer (see Fig.\
\ref{fig:setup}).
We neglect fluctuations of the interaction energy $U$, which is
less susceptible to the electrostatic environment of the impurities
than $\varepsilon$.

We assume that the density of impurities is small enough, 
\begin{equation}
  \hbar \rho d^3 \Gamma \ll 1,
  \label{eq:singleimpurity}
\end{equation}
where $\hbar \Gamma$ is the typical width of
an impurity energy level, so that the transport of an electron will
take place through a single impurity only, {\em i.e.}, multiple-impurity
processes are ignored. The total current $I$ and torque $T$ are then
written as integrals over the energy $\varepsilon$ of impurities in
the junction,
\begin{equation}
  I = \int_{\mu_L}^{\mu_R} d\varepsilon\,
  [ i'(\varepsilon) + i''(\varepsilon)],\ \
  T = \int_{\mu_L}^{\mu_R} d\varepsilon\,[
    t'(\varepsilon) + t''(\varepsilon)].
\end{equation}
where $i'$, $t'$ and $i''$, $t''$
represent contributions from impurities at energy
$\varepsilon$ (case I) and $\varepsilon - U$ (case II),
respectively.

For barriers with rough interfaces the transverse momentum is
not be conserved at the ferromagnet-insulator interface. In that case,
the tunneling rates $\Gamma_{\alpha \sigma}$ can be estimated
as \cite{kn:slonczewski2005}
\begin{equation}
  \Gamma_{\alpha\sigma} = P_{\alpha\sigma} e^{-2 x/\lambda},
  \label{eq:GammaEst}
\end{equation}
where $P_{\alpha\sigma}$ is a polarization factor and $\lambda$ the
wavefunction decay length in the spacer layer. \cite{foot}
 We will take the junction to be symmetric, $P_{L\sigma} = P_{R\sigma}=P_\sigma$, and
define $P_{\mp}= (P_\uparrow \mp P_\downarrow)/2$. For barrier width
$d \gg \lambda$ the current and torque are dominated by impurities
near the center of the barrier, $x
\approx d/2$. \cite{kn:manchon2007}
The contribution of these impurities to the current
and torque scales
$\sim e^{-d/\lambda}$ if $d \gg \lambda$, whereas there is a faster
exponential suppression for direct tunneling or for impurities located
off-center.

%

\begin{figure}
\centering{}
\includegraphics[width=0.95\hsize]{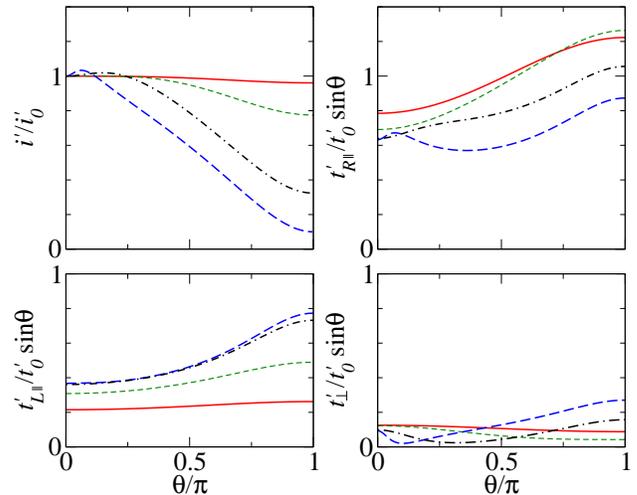}
\caption{Spectral current $i'(\varepsilon)$ and torque components
  $t'_{R\parallel}(\varepsilon)$, $t'_{L\parallel}(\varepsilon)$, and
  $t'_{\perp}(\varepsilon)$ versus the angle $\theta$ between the
  magnetizations of the left and right ferromagnetic contacts.
The curves shown are for polarization $p  =
  0.2$ (solid), $0.5$ (short dash), $0.9$ (dash-dot), and $0.99$
  (long dash). For definiteness, the exchange field parameters have been
  set to
  $\eta_L=\eta_R=1$.\label{fig:2}}
\end{figure}

We have calculated the two contributions to the current and torque for
arbitrary angle and polarization $p = P_-/P_+$. The results are rather
lengthy, and we refer to Fig.\ \ref{fig:2} for a numerical evaluation
for a few representative values of $p$. Closed form
expressions could be obtained in the limit of weakly polarized
ferromagnets, $p \ll 1$ only. For impurities with
$\mu_R < \varepsilon < \mu_L$ (case I), we find
\begin{eqnarray}
  i' &=&
  i_0' \left(1 + \frac{1}{2} p^2 \cos \theta
  + \frac{1}{2} p^{5/2} \eta_L \sin^2 \theta + \ldots \right),
  \nonumber \\
  t'_{R\parallel} &=&
  t_0' \sin \theta
  \left( 1 - \frac{1}{2} p^{1/2} \eta_L \cos \theta + \ldots \right), \nonumber \\
  t'_{L\parallel} &=& -
  \frac{t_0'}{4} \eta_L \sin \theta
  \left( 2 p^{1/2} - p^{3/2}  \eta_R^2 \cos \theta
  + \ldots
  \right),  \nonumber \\
  t_{\perp}' &=& \frac{t_0'}{4} \eta_L \sin \theta
  \left( 2 p^{1/2}  + p^{3/2}  \eta_R^2  \cos \theta
  + \ldots \right),~~~~~
  \label{eq:ittt}
\end{eqnarray}
where
\begin{equation}
  i_0'(\varepsilon) = \pi 2^{-7/2} P_+ \lambda \rho(\varepsilon,d/2)
e^{-d/\lambda},\ \ t_0'(\varepsilon) = 2\hbar p i_0'(\varepsilon),
\end{equation}
and $\eta_{\alpha} = \sqrt{|h_{\alpha}|/\Gamma_{\alpha-}}$. In Eq.\
(\ref{eq:ittt}) we kept only those sub-leading terms in the
small-$p$ expansion that have a different $\theta$ dependence than
the leading terms. The case II contributions are obtained by
interchanging $L$ and $R$ and replacing $\rho(\varepsilon,d/2)$ by
$\rho(\varepsilon-U,d/2)$. The total torque is found by adding
contributions for cases I and II and integrating over $\varepsilon$.

There is a striking asymmetry between the impurity-mediated
in-plane torques on the source and drain reservoirs. For small
polarization ($p \ll 1$) and case I, the in-plane torque
$t_{R\parallel}$ dominates over the other two torque terms. This
is markedly different from the torque from direct
tunneling, which has equal magnitudes for $t_{L\parallel}$ and
$t_{R\parallel}$. \cite{kn:slonczewski2005} The situation is
reversed for case II. However,
as there is no a priori reason why the spectral impurity
densities at energies $\varepsilon$ and $\varepsilon-U$, which
set the sizes of the torques for cases I and II, are equal, one
still expects the magnitudes $t_{L\parallel}$ and $t_{R\parallel}$ to
be rather different after torques from the two cases are added.

It is important to point out that both the source-drain asymmetry
and the existence of an out-of-plane component of the torque are
interaction effects. The source-drain asymmetry was already discussed 
at the end of Sec.\ \ref{sec:2}. That the out-of-plane torque is an 
interaction effect can be seen explicitly from Eq.\ (\ref{eq:tp}), 
which is proportional to $h_L$ ($h_R$ for case II). Without interactions,
$h_L = h_R = 0$. (Note that the out-of-plane torque in theories
without electron-electron interactions has an altogether different
origin: It is caused by interference effects.\cite{kn:manchon2007,kn:manchon2008,kn:xiao2008}
 For the temperature range we consider, $T \gg \hbar \Gamma$, interference effects are smeared out
for impurity-mediated transport.)


For strongly polarized ferromagnetic contacts ($p \sim 1$) all three
torque terms are of comparable magnitude, although $t_{L\parallel}$
and $t_{R\parallel}$ remain different. The order of magnitude of
the torque, normalized to the impurity-mediated current $i'$, is the
same as for the case of direct tunneling, normalized to the direct 
tunneling current \cite{kn:slonczewski2005}.
Whereas the angular dependence of all three torque terms is the
standard geometric $\sin \theta$ dependence characteristic of the
direct tunneling torque
\cite{kn:slonczewski2005,kn:theodonis2006,kn:kalitsov2006,kn:xiao2008,kn:manchon2008}
if $p \ll 1$, the $\theta$ dependence is more complicated for strongly
polarized contacts because of the presence of the exchange fields $h_L$
and $h_R$, see Fig.\ \ref{fig:2}.



\section{Discussion}

The impurity-mediated torque considered here coexists with the torque 
from direct tunneling. How can the two contributions to the torque be 
separated experimentally? The calculation of the previous sections shows
that there are four significant differences between the impurity-mediated 
torque and the direct-tunneling torque: (i) The impurity-mediated torque 
is not symmetric under reversal of the bias, whereas the torque from 
direct tunneling is symmetric under bias reversal. The same holds for
the impurity-mediated and direct-tunneling currents. Hence, the presence
of a bias asymmetry for the current or torque is an indicator for the
order-of-magnitude of the impurity-mediated contribution.
(ii) The impurity-mediated torque has a nontrivial angular dependence 
for strongly polarized contacts, whereas the direct-tunneling torque has
a sinusoidal angular dependence. The angular dependence of the torque is 
observable in current-induced ferromagnetic resonance 
experiments.\cite{kn:tulapurkar2005,kn:sankey2008} 
(iii) Impurity-mediated and direct-tunneling contributions to the 
current and torque have different dependences on the thickness $d$ of the 
insulating layer: The impurity-mediated contribution scales
$\propto e^{-d/\lambda}$, whereas the direct-tunneling contribution
scales $\propto e^{-2 d/\lambda}$. This means that the impurity-mediated 
torque will dominate for sufficiently thick junctions, irrespective 
of the impurity concentration. 
(iv) The bias-dependence of the impurity-mediated torque tracks the 
spectral impurity density $\rho(\varepsilon,x)$, whereas the 
bias-dependence of the direct-tunneling torque should be less
pronounced. (The spectral impurity density is not
the only source of a bias dependence: One also expects a
bias dependence from the bias-dependence of the tunneling rates
$\Gamma$. However, that bias dependence is likely to be the shared
between the two contributions to the torque.) For the impurity-mediated torque,
 resonant features in the spectral density, which can be brought 
out using more detailed theoretical
modeling \cite{kn:tsymbal2003,kn:velev2007b,kn:zhuravlev2005} or
additional experiments,\cite{kn:miao2008,kn:jansen1999} can then 
be correlated with the bias dependence of the torque. 

It is instructive to give a rough comparison of the magnitudes of 
the impurity-mediated and direct tunneling currents or torques. 
Since they have different exponential dependences on the barrier 
thickness $d$, we estimate the pre-exponential factors by order
 of magnitude only and neglect differences between the three relevant 
microscopic length scales (Fermi wavelengths for majority and minority 
electrons in the ferromagnetic contacts, wavefunction decay length 
$\lambda$). Dimensional analysis then estimates the ratio 
of impurity-mediated and direct currents $I_{\rm impurity}$, $I_{\rm direct}$ and torques $T_{\rm impurity}$, $T_{\rm direct}$ as
\begin{equation}
  I_{\rm impurity}/I_{\rm direct} \sim T_{\rm impurity}/T_{\rm direct}
  \sim 
  e^{d/\lambda} n \lambda^3,
\end{equation}
where $n = \int d\varepsilon \rho(\varepsilon)$ is the impurity 
concentration. Reference \onlinecite{kn:zhuravlev2005} uses 
$n \sim 10^{27}$ m$^{-3}$, whereas an estimate using the barrier height in MgO and the effective mass of the electron gives\cite{kn:faure-vincent2002} 
$\lambda \sim 10^{-9}$ m. From this, we conclude that impurity-mediated transport dominates already for barrier thickness $d \gtrsim 10^{-9}$ m, well within the experimentally relevant range. A more accurate comparison requires knowledge about the spectral density specific to the impurity type and is beyond the scope of this article.

\begin{acknowledgments}
We acknowledge helpful discussions with Bob Buhrman, Dan Ralph, John
Read, Joern Kupferschmidt and Lin Xue. This work was supported by
the Cornell Center for Nanoscale Systems under NSF Grant No.\
EEC-0117770 and by the NSF under Grant No.\ DMR 0705476.
\end{acknowledgments}


\end{document}